\documentclass[aps]{revtex4}
\usepackage{amssymb,epsf}

\begin{document}

\title{Exact solutions of three dimensional black holes:\\
Einstein gravity vs $F(R)$ gravity}
\author{S. H. Hendi$^{1,2}$, B. Eslam Panah$^{1}$ and R. Saffari$^{3}$}
\affiliation{$^1$ Physics Department and Biruni Observatory, College of Sciences, Shiraz
University, Shiraz 71454, Iran\\
$^2$ Research Institute for Astronomy and Astrophysics of Maragha (RIAAM),
Maragha, Iran\\
$^3$ Department of Physics, University of Guilan, P. O. Box: 41335-1914,
Rasht, Iran}

\begin{abstract}
In this paper, we consider Einstein gravity in the presence of a class of
nonlinear electrodynamics, called power Maxwell invariant (PMI). We take
into account $(2+1)$-dimensional spacetime in Einstein-PMI gravity and
obtain its black hole solutions. Then, we regard pure $F(R)$ gravity as well
as $F(R)$-conformally invariant Maxwell theory to obtain exact solutions of
the field equations with black hole interpretation. Finally, we investigate
the conserved and thermodynamic quantities and discuss about the first law
of thermodynamics for the mentioned gravitational models.
\end{abstract}

\maketitle

\section{introduction}

In recent years, the luminosity distance of Supernovae type Ia \cite%
{Perlmutter}, wide surveys on galaxies \cite{Gal} and the
anisotropy of cosmic microwave background radiation \cite{CMBR}
confirm that the expansion of our Universe is currently undergoing
a period of acceleration. Large scale structure formation
\cite{LSS}, baryon oscillations \cite{BO} and weak lensing
\cite{WL} also suggest such an accelerated expansion of the
Universe. Identifying the cause of this late time acceleration is
one of the most challenging problems of modern cosmology.
Theoretical physicists desire to interpret this accelerated
expansion in a suitable gravitational background and they proposed
some candidates. A positive cosmological constant can lead to
accelerated expansion of the universe but it is plagued by the
fine tuning problem \cite{Copeland}. The cosmological constant may
be interpreted either geometrically as modifying the left hand
side of Einstein's equation or as a kinematic term on the right
hand side with the equation of state parameter $w = -1$. Another
approach can further be generalized by considering a source term
with an equation of state parameter $w < -1/3$. Such kinds of
source terms have collectively come to be known as Dark Energy.
Various scalar field models of dark energy have been considered in
literature \cite{Armendariz}. All the dark energy based theories
assume that the observed acceleration is the outcome of the action
of a still unknown ingredient added to the cosmic pie. In terms of
the Einstein equations, such models are simply modifying the right
hand side including in the stress--energy tensor with something
more than the usual matter and radiation components.

On the other hand, one can also try to leave unchanged the source side, but
rather than modifying the left hand side of Einstein field equations. In a
sense, one is therefore interpreting cosmic acceleration as a first signal
of the breakdown of the laws of physics as described by the standard General
Relativity (GR). There are different branches of modified gravity with
various motivations. Lovelock gravity \cite{Lovelock}, brane world cosmology
\cite{Gergely}, scalar-tensor theories \cite{Jordan} and also the so-called $%
F(R)$ gravity \cite{FR,HendiGRG,HendiPRD} are some of modified gravity
theories.

Modifying GR, not simply given its positive results, opens the way to a
large class of alternative theories of gravity ranging from extra dimensions
\cite{Dvali} to non-minimally coupled scalar fields \cite{Caresia}. In
particular, we will be interested here in fourth order theories \cite%
{Capozziello} based on replacing the scalar curvature R in the
Hilbert--Einstein action with a generic analytic function F(R) which should
be reconstructed starting from data and physically motivated issues.

In this paper we are interested in $F(R)$ gravity. But as we know the field
equations of $F(R)$ gravity are complicated fourth-order differential
equations, and it is not easy to find exact analytical solutions. In
addition, adding stress-tensor of a matter field to $F(R)$ gravity, increase
its difficulties. Recently, it has been shown that one can extract exact
analytical solutions of $F(R)$ theory coupled to a traceless energy momentum
tensor with constant curvature scalar \cite{Moon}. For example, taking into
account the conformally invariant Maxwell (CIM) field as a matter source,
which is traceless in arbitrary dimensions, some black objects of $F(R)$
gravity were obtained in higher dimensions \cite{Sheykhi}.

On the other hand, one of the interesting subjects for recent study is the
investigation of three dimensional black holes \cite{Hodgkinson}. Taking
into account three dimensional solutions helps us to find a profound insight
in the black hole physics, quantum view of gravity and also its relations to
string theory \cite{Carlip,Witten}. Moreover, three dimensional spacetimes
perform an essential role to improve our understanding of gravitational
interaction in low dimensional manifolds \cite{Witten2007}. Due to these
facts, some of physicists have an interest in the $(2+1)$-dimensional
manifolds and their attractive properties \cite{BTZ,Nojiri1998}. Although
three dimensional black holes in $F(R)$ gravity have been studied before
\cite{ThreeF(R),HendiIJPT}, till now, exact solution of three dimensional $%
F(R)$ gravity coupled to a matter field have not been constructed. In this
paper, one of our goals is obtaining an exact three dimensional black hole
solutions of $F(R)$ theory coupled to a CIM source.

The coupling of nonlinear sources and general relativity attract the
significant attentions because of their specific properties. Interesting
properties of various nonlinear electrodynamics have been studied before
\cite{NLED}. One of the special class of the nonlinear electrodynamics
sources is PMI, which its Lagrangian is an arbitrary power of Maxwell
Lagrangian \cite{Hassaine2008}. This Model is considerably richer than
Maxwell theory and in the special case (unit power), it reduces to linear
Maxwell field. Another attractive feature of the PMI theory is its conformal
invariance when the power of Maxwell invariant is a quarter of spacetime
dimensions. In other words, for the special choice $power=dimensions/4$, one
obtains traceless energy-momentum tensor which leads to conformal
invariance. It is notable that the idea is to take advantage of the
conformal symmetry to construct the analogues of the four dimensional
Reissner-Nordstr\"{o}m solutions with an inverse square electric field in
arbitrary dimensions \cite{Hassaine2007}.

Recently it has been shown that one can, simultaneously, extract electric
charge and cosmological constant from pure $F(R)$ gravity (without matter
field: $T_{\mu \nu }=0$) \cite{HendiGRG}. Another goal of this paper is
obtaining three dimensional charged black hole solutions from pure $F(R)$
gravity as well as $F(R)$-CIM gravity and compare them.

The outline of our paper is as follows. In Sec. II, we review three
dimensional black hole solutions in Einstein-Maxwell gravity. Then we
investigate the black hole solutions of Einstein-PMI and Einstein-CIM
theories. Sec. III is devoted to obtain black hole solutions of pure $F(R)$
gravity as well as $F(R)$-CIM theory and compare these solutions. In Sec.
IV, we discuss about the conserved and thermodynamic quantities of the
solutions and check the first law of thermodynamics. We terminate our paper
by some conclusions.

%%%%%%%%%%%%%%%%%%%%%%%%%%%%%%%%%%%%%%%%%%%%%%%%%%%%%%%%%%%%%%%%%%%%%%%%%%%%%%%%%%%%%%%%%%%%%%%%%

\section{Three dimensional solutions in Einstein gravity}

\subsection{\emph{Brief review of Einstein-Maxwell solutions}}

The charged BTZ black hole is the solution of the $(2+1)$-dimensional
Einstein-Maxwell gravity with a negative cosmological constant $\Lambda =-%
\frac{1}{l^{2}}$ \cite{BTZ2000}. The line element can be written as
\begin{equation}
ds^{2}=-g(r)dt^{2}+\frac{dr^{2}}{g(r)}+r^{2}d\varphi ^{2},  \label{Metric}
\end{equation}%
where the metric function is
\begin{equation}
g(r)=\frac{r^{2}}{l^{2}}-m-2q^{2}\ln (\frac{r}{l}),  \label{g(r)BTZ}
\end{equation}%
where $m$ and $q$ are the mass and the electric charge of the black hole,
respectively.

\begin{figure}[tbp]
\epsfxsize=8cm \centerline{\epsffile{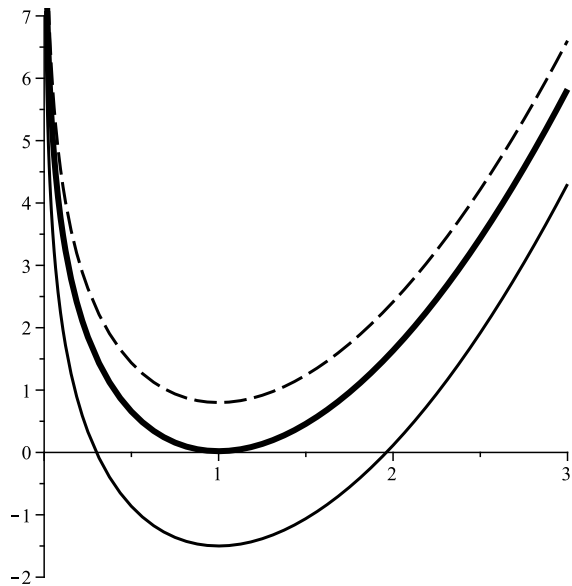}} \caption{$g(r)$
(Eq. (\protect\ref{g(r)BTZ})) versus $r$ for $l =1$, $q=1$, and
$m=0.2$ (dashed line), $m_{ext}=0.98$ (bold line) and $m=2.5$
(continuous line).} \label{Fig1}
\end{figure}

Here, we want to review the geometrical structure of this solution, briefly.
We first look for the essential singularity(ies). The Ricci scalar and the
Kretschmann scalar can be written in the following form
\begin{eqnarray}
R &=&\frac{2\left( q^{2}l^{2}-3r^{2}\right) }{r^{2}l^{2}},
\label{RiccichrBTZ} \\
R_{\alpha \beta \mu \nu }R^{\alpha \beta \mu \nu } &=&\frac{4\left(
3r^{4}-2r^{2}q^{2}l^{2}+3q^{4}l^{4}\right) }{r^{4}l^{4}},  \label{KrechrBTZ}
\end{eqnarray}%
which indicate that
\begin{eqnarray}
\lim_{r\longrightarrow 0}R &\longrightarrow &\infty  \nonumber \\
\lim_{r\longrightarrow 0}R_{\alpha \beta \gamma \delta }R^{\alpha \beta
\gamma \delta } &\longrightarrow &\infty ,
\end{eqnarray}%
and so confirm that there is a curvature singularity at $r=0$. Also The
Ricci and Kretschmann scalars are $\frac{-6}{l^{2}}$ and $\frac{12}{l^{4}}$
at $r\longrightarrow \infty $, and one concludes \ that the asymptotic
behavior of the charged BTZ black hole is adS. Also we plot $g(r)$ versus $r$
in Fig. \ref{Fig1}, to show that the solution (\ref{g(r)BTZ}) may be
interpreted as naked singularity or black hole with two horizons or extreme
black hole.
%%%%%%%%%%%%%%%%%%%%%%%%%%%%%%%%%%%%%%%%%%%%%%%%%%%%%%%%%%%%%%%%%%%%%%%%%%%%%%%%%%%%%%%%%%%%%%%%%
%%%%%%%%%%%%%%%%%%%%%%%%%%%%%%%%%%%%%%%%%%%%%%%%%%%%%%%%%%%%%%%%%%%%%%%%%%%%%%%%%%%%%%%%%%%%%%%%%%

\subsection{ \emph{Einstein-PMI and Einstein-CIM solutions}}

Now, we take into account the Einstein gravity in the presence of a matter
source in the form $\left( F_{\mu \nu }F^{\mu \nu }\right) ^{s}$, so called
Einstein-PMI gravity \cite{Hassaine2008}. Black hole solutions in $(n+1)$%
-dimension of Einstein-PMI gravity have been studied before \cite%
{HendiEPJC2011} for spacial value of $s$. Here, we want to focus on the $%
(2+1)$-dimension of Einstein-PMI gravity for arbitrary $s$ and discuss about
the properties of the solutions. The $(2+1)$-dimensional action in which
gravity is coupled to nonlinear electrodynamics in the presence of negative
cosmological constant may be written as \cite{HendiEPJC2011}
\begin{equation}
I~\left( g_{\mu \nu },~A_{\mu }\right) =\frac{1}{16\pi }\int_{\partial
M}d^{3}x\sqrt{-g}\left[ R-2\Lambda +\left( \kappa F\right) ^{s}\right] ,
\label{ActionPM}
\end{equation}
where $R$ is the scalar curvature, $F$ is the Maxwell invariant which is
equal to $F_{\mu \nu }F^{\mu \nu }$ (where $F_{\mu \nu }=\partial _{\mu
}A_{\nu }-\partial _{\nu }A_{\mu }$ is the electromagnetic tensor field and $%
A_{\mu }$ is the gauge potential), and $s$ is an arbitrary positive
nonlinearity parameter ($s\neq \frac{1}{2}$). Varying the action (\ref%
{ActionPM}) with respect to the metric tensor $g_{\mu \nu }$ and the
electromagnetic field $A_{\mu }$, the equations of gravitational and
electromagnetic fields may be obtained as
\begin{equation}
G_{\mu \nu }-\Lambda g_{\mu \nu }=T_{\mu \nu },  \label{Eq1PM}
\end{equation}%
\begin{equation}
\partial _{\mu }\left( \sqrt{-g}F^{\mu \nu }\left( \kappa F\right)
^{s-1}\right) =0,  \label{Eq2PM}
\end{equation}%
in which the energy-momentum tensor of Eq. (\ref{Eq1PM}) is
\begin{equation}
T_{\mu \nu }=2\left[ s\kappa F_{\mu \rho }F_{\nu }^{\rho }\left( \kappa
F\right) ^{s-1}-\frac{1}{4}g_{\mu \nu }\left( \kappa F\right) ^{s}\right] ,
\label{TPM}
\end{equation}%
where $\kappa $ is a constant. It is easy to show that when $s$ and $\kappa$
go to $-1$, Eqs. (\ref{ActionPM}-\ref{TPM}), reduce to the field equations
of black hole in Einstein-Maxwell gravity. Since the Maxwell invariant is
negative, hereafter we set $\kappa =-1$, without loss of generality.

We look for black hole solutions with a radial electric field, so the gauge
potential is given by
\begin{equation}
A_{\mu }=h(r)\delta _{\mu }^{0},  \label{gauge}
\end{equation}%
where the electromagnetic field equation (\ref{Eq2PM}), leads to
\begin{equation}
h(r)=\left\{
\begin{array}{c}
q\ln (\frac{r}{l})\text{ \ \ \ \ \ \ \ }s=1 \\
-qr^{\left( \frac{2(s-1)}{2s-1}\right) } \text{ \ \ \ \ \ \ \
}otherwise
\end{array}%
\right. ,  \label{h(r)}
\end{equation}

where $q$ is an integration constant related to electric charge.
It is easy to show that the only nonzero electromagnetic field
tensor is
\begin{equation}
F_{tr}=\left\{
\begin{array}{c}
\frac{q}{r}\text{ \ \ \ \ \ \ \ \ \ \ \ \ }s=1 \\
\frac{-2q\left( s-1\right) }{(2s-1)}r^{\frac{-1}{2s-1}}\text{ \ \ }otherwise%
\end{array}%
\right. .  \label{Ftr}
\end{equation}

In order to have asymptotically well-defined electric field, we should
restrict the nonlinearity parameter to $s>\frac{1}{2}$.

To find the function $g(r)$, one may use the components of Eq. (\ref{Eq1PM}%
). The simplest equation is the $rr$ (or $tt$) component of this equation
which can be written as
\begin{equation}
\left\{
\begin{array}{c}
g^{\prime }(r)+2\Lambda r-\frac{2q^{2}}{r}=0\text{ \ \ \ \ \ \ \ \
\ \ \ \ \ \ \ \ \ \
\ \ \ \ \ \ \ \ \ \ \ \ \ \ \ \ \ \ \ }s=1 \\
g^{\prime }(r)+2\Lambda r+r\left( 1-2s\right) \left( \frac{8q^{2}r^{\left(
\frac{-2}{2s-1}\right) }\left( s-1\right) ^{2}}{\left( 2s-1\right) ^{2}}%
\right) ^{s}=0\text{ \ \ }otherwise%
\end{array}
,\right.  \label{EqS1}
\end{equation}
with the following solutions
\begin{equation}
g(r)=\frac{r^{2}}{l^{2}}-m+\left\{
\begin{array}{c}
2q^{2}\ln (\frac{r}{l})\text{ \ \ \ \ \ \ \ \ \ \ \ }s=1 \\
\frac{\left( 2s-1\right) ^{2}\left( \frac{8q^{2}\left( s-1\right) ^{2}}{%
\left( 2s-1\right) ^{2}}\right) ^{s}}{2\left( s-1\right) }r^{\left( \frac{%
2\left( s-1\right) }{2s-1}\right) }\text{ \ }otherwise%
\end{array}
,\right.  \label{g(r)PMI}
\end{equation}
where $m$ is an integration constant related to mass. We should
note that Eq. (\ref{g(r)PMI}) satisfy all field equations. We
should note that since $s>\frac{1}{2}$, the first term of Eq. (\ref{g(r)PMI}%
) is dominant term for $r\rightarrow \infty $ and therefore the asymptotic
behavior of the solutions is adS. In other word, the nonlinearity does not
affect on the asymptotic behavior of the solutions.

Now, we want to investigate the special case $s=\frac{3}{4}$, the so-called
conformally invariant Maxwell field. It has been shown that for $s=\frac{d}{4%
}$ ($d$=spacetime dimension), the energy-momentum tensor will be traceless
and the corresponding electric field will be proportional to $r^{-2}$ as it
take place for Maxwell field in four dimension. Here, we consider $s=\frac{3%
}{4}$ (see for more details \cite{Hassaine2007}) into Eqs. (\ref{Ftr}) and (%
\ref{g(r)PMI}) to obtain%
\begin{equation}
F_{tr}=\frac{q}{r^{2}},
\end{equation}%
\begin{equation}
g(r)=\frac{r^{2}}{l^{2}}-m-\frac{\left( 2q^{2}\right) ^{\frac{3}{4}}}{2r}.
\label{g(r)CIM}
\end{equation}%
\ \ \

We should note that for $PMI$ solution, the metric function $g(r)$ has one
real root (such as uncharged solutions). This behavior take place for the
metric function (\ref{g(r)PMI}), when we choose $s\neq 1$ (see Fig. \ref%
{Fig2}).
%%%%%%%%%%%%%%%%%%%%%%%%%%%%%%%%%%%%%%%%%%%%%%%%%%%%%%%%%%%%%%%%%%%%%%%%%%%%%%%%%%%%%%%%%%%%%%%%%
\begin{figure}[tbp]
\epsfxsize=8cm \centerline{\epsffile{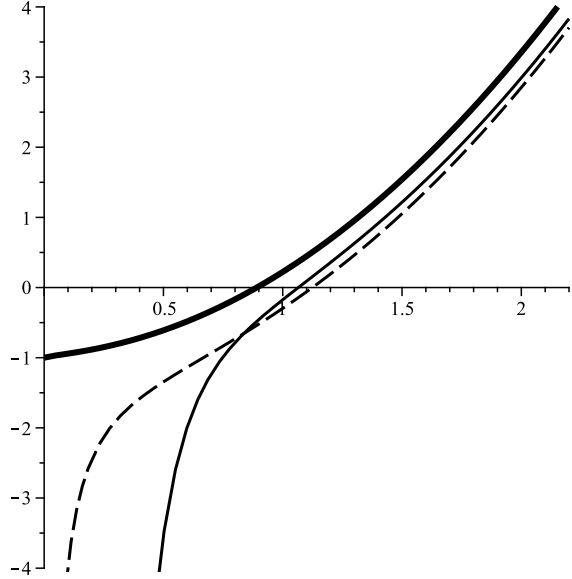}}
\caption{$g(r)$\ (Eq. (14)) versus $r$ for $\l =1$, $q=0.5$, $m=1$ and $%
s=3/4 $ (dashed line), $s=2$ (bold line) and $s=3/5$ (continuous line).}
\label{Fig2}
\end{figure}
%%%%%%%%%%%%%%%%%%%%%%%%%%%%%%%%%%%%%%%%%%%%%%%%%%%%%%%%%%%%%%%%%%%%%%%%%%%%%%%%%%%%%%%%%%%%%%%%%

\section{\textbf{Three dimension solutions in F(R) gravity}}

\subsection{\textit{F(R)-CIM solution}}

In this section, we consider $F(R)$ gravity in the presence the
conformally invariant Maxwell field as a source, which leads to
traceless energy-momentum tensor.

The equations of motion of $F(R)-CIM$ theory can be written as
\begin{equation}
R_{\mu \nu }\left( 1+f_{R}\right) -\frac{1}{2}g_{\mu \nu }F(R)+\left( g_{\mu
\nu }\nabla ^{2}-\nabla _{\mu }\nabla _{\nu }\right) f_{R}=8\pi \mathrm{T}%
_{\mu \nu },  \label{EqF(R)1}
\end{equation}%
\begin{equation}
\partial _{\mu }\left( \sqrt{-g}\frac{F^{\mu \nu }}{\left( -F\right) ^{\frac{%
1}{4}}}\right) =0,  \label{EqF(R)2}
\end{equation}%
where $f_{R}=\frac{df(R)}{dR}$. It is notable that the assumption of a
traceless energy-momentum tensor is essential for deriving exact black hole
solutions in $f(R)$ gravity coupled to the matter field in metric formalism.

Now, we want to obtain the solutions for the constant scalar curvature is $%
R=R_{0}$= const. Using Eq. (\ref{gauge}) with (\ref{EqF(R)2}), we can obtain%
\begin{equation}
h(r)=\frac{-\mathcal{Q}}{r},
\end{equation}%
\begin{equation}
F_{tr}=\frac{\mathcal{Q}}{r^{2}},
\end{equation}

The trace of Eq. (\ref{EqF(R)1}) yields%
\begin{equation}
R_{0}\left( 1+f_{R}\right) -\frac{3}{2}\left( R_{0}+f(R_{0})\right) =0,
\end{equation}%
and solving the above equation for $R_{0}$ gives%
\begin{equation}
R_{0}=\frac{3f(R_{0})}{2f_{R}-1}\equiv 6\Lambda .
\end{equation}

Substituting the above relation into Eq. (\ref{EqF(R)1}), we obtain the
following equation%
\begin{equation}
R_{\mu \nu }\left( 1+f_{R}\right) -\frac{1}{3}g_{\mu \nu }R_{0}\left(
1+f_{R}\right) =8\pi \mathrm{T}_{\mu \nu }.  \label{Filedeq1}
\end{equation}

Considering metric (\ref{Metric}) and field equation (\ref{Filedeq1}), one
can write the following field equations%
\begin{equation}
\left[ r^{2}\left( 2\mathcal{Q}^{2}\right) ^{\frac{1}{4}}\left( rg^{\prime
\prime }(r)+g^{\prime }(r)+\frac{2}{3}rR_{0}\right) \right] \left(
1+f_{R}\right) =-\mathcal{Q}^{2},  \label{e1}
\end{equation}%
\begin{equation}
\left[ 3g^{\prime }(r)+rR_{0}\right] \left( 1+f_{R}\right) =\frac{3}{2r^{2}}%
\left( 2\mathcal{Q}^{2}\right) ^{\frac{3}{4}},  \label{e2}
\end{equation}%
corresponding to $tt\ $(or $rr$) and $\varphi \varphi $\ components,
respectively. After some calculations, one can obtain the following metric
function%
\begin{equation}
g(r)=-m-\frac{\left( 2\mathcal{Q}^{2}\right)
^{\frac{3}{4}}}{2\left( 1+f_{R}\right) r}-\frac{r^{2}R_{0}}{6},
\label{g(r)F(R)}
\end{equation}%
where we restrict ourselves to $f^{\prime }(R_{0})\neq -1$. We want to study
the general structure of the solutions. One can find that the Kretschmann
scalar is%
\begin{equation}
R_{\alpha \beta \gamma \delta }R^{\alpha \beta \gamma \delta }=\frac{%
2r^{6}R_{0}^{2}\left( 1+f_{R}\right) ^{2}+9(2\mathcal{Q}^{2})^{\frac{3}{2}}}{%
6r^{6}\left( 1+f_{R}\right) ^{2}},  \label{KreshF(R)-CIM}
\end{equation}
where confirms that the Kretschmann scalar diverges at $r=0$, is
finite for $ r\neq 0$, and is proportional to
$\frac{R_{0}^{2}}{3}$ as $r\longrightarrow \infty $. Therefore,
there is a curvature singularity located at $r=0$ and the
spacetime is asymptotically adS provided we define
$R_{0}^{{}}=-\frac{6}{ l^{2}}$. In order to have physical
solutions we should restrict ourselves to $f_{R} \neq -1$. For $
f_{R}>-1$, one can substitute $R_{0}$ and $\mathcal{Q}$ with
$-\frac{6}{ l^{2}}$ and $q[1+f_{R}]^{\frac{2}{3}}$, respectively
to obtain the metric function of Einstein-$CIM$ gravity
(\ref{g(r)CIM}) with one non-extreme root (such as Schwarzschild
black hole).

\subsection{\emph{Pure $F(R)$ gravity solutions}}

Here, we are looking for charged solutions of pure $F(R)$ gravity. We
consider $F(R)=R+f(R)$ gravity without matter field ($\mathrm{T}_{\mu \nu
}^{matt}=0$) in $(2+1)$-dimension.

Considering metric (\ref{Metric}) and field equation (\ref{EqF(R)1}), one
can obtain the following independent sourceless equations \cite{HendiIJPT}%
\begin{eqnarray}
&&2rg(r)F_{R}^{\prime \prime }+\left[ rg^{\prime }(r)+2g(r)\right]
F_{R}^{\prime }-\left[ rg^{\prime \prime }(r)+g^{\prime }(r)\right]
F_{R}=rF(R),  \label{Fieldtt} \\
&&\left[ rg^{\prime }(r)+2g(r)\right] F_{R}^{\prime }-\left[ rg^{\prime
\prime }(r)+g^{\prime }(r)\right] F_{R}=rF(R),  \label{Fieldrr} \\
&&2rg(r)F_{R}^{\prime \prime }+2g^{\prime }(r)\left[ rF_{R}^{\prime }-F_{R}%
\right] =rF(R),  \label{Fieldphiphi}
\end{eqnarray}%
corresponding to $tt$, $rr$ and $\varphi \varphi $ components, respectively.
In these equation $F_{R}=\frac{dF(R)}{dR}$ and prime denotes derivative with
respect to $r$. In this paper, we study black hole solutions with constant
Ricci scalar (so $F_{R}^{\prime \prime }=F_{R}^{\prime }=0$) , therefore it
is easy to show that the field Eqs. (\ref{Fieldtt})-(\ref{Fieldphiphi})
reduce to
\begin{eqnarray}
\left[ rg^{\prime \prime }(r)+g^{\prime }(r)\right] F_{R} &=&-rF(R),
\label{eqaa} \\
2g^{\prime }(r)F_{R} &=&-rF(R).  \label{eqbb}
\end{eqnarray}

Hereafter, we should choose a model of $f(R)$ to obtain exact solution. To
find the function $g(r)$, we use the components of Eqs . (\ref{eqaa}) and (%
\ref{eqbb}) with well-known $F(R)=R-\lambda \exp (-\xi R)+\eta R^{n}$ model.
Using of Eqs. (\ref{eqaa}) and (\ref{eqbb}) with metric (\ref{Metric}), one
can obtain a general solution in the following form
\begin{equation}
g(r)=\frac{r^{2}}{l^{2}}-m+\frac{K}{r},  \label{g(r)pure}
\end{equation}
where $K$ is integration constants. In order to satisfy all components of
the field equation, we should set the parameters of the $F(R)$ model such
that satisfy the following equations
\begin{equation}
R\left( \eta (2n-3)R^{n-1}-1\right) e^{\xi R}+\lambda \left( 3+2\xi R\right)
=0,  \label{FR1}
\end{equation}%
\begin{equation}
\eta nR\left( R^{n-1}-1\right) e^{\xi R}-6\lambda \xi =0.  \label{FR2}
\end{equation}

Solving Eqs. (\ref{FR1}) and (\ref{FR2}) for $\lambda $ and $\eta $ with
arbitrary $\xi $ lead to the following solutions
\begin{equation}
\eta =\frac{-\left( 1+\xi R\right) }{R^{n-1}\left( n+\xi R\right) },
\label{lambda1}
\end{equation}%
\begin{equation}
\lambda =\frac{6e^{\xi R}\left( n-1\right) }{n+\xi R}.  \label{kappa1}
\end{equation}

\begin{figure}[tbp]
\epsfxsize=8cm \centerline{\epsffile{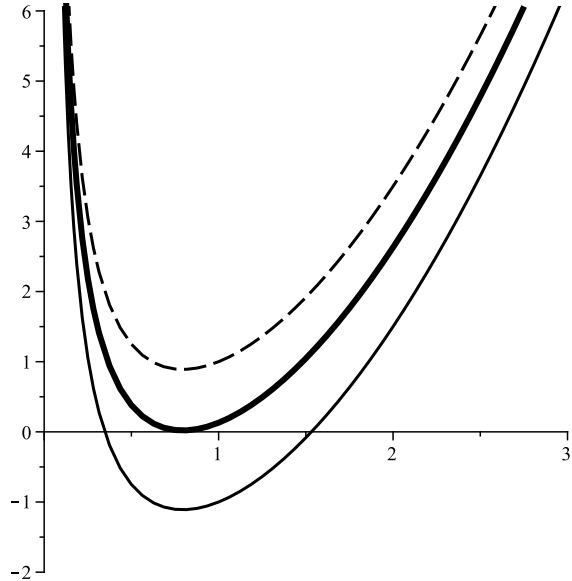}}
\caption{$g(r)$ (Eq. (\ref{g(r)pure})) versus $r$ for $\l =1$, $K=1$, and $m=1$ (dashed line), $%
m=1.87$ (bold line) and $m=3$ (continuous line).} \label{Fig3}
\end{figure}

The solution (\ref{g(r)pure}) known as uncharged BTZ black hole solution,
when $K=0$ \cite{BTZ}. Calculations show that the Kretschmann scalar is in
the following form
\begin{equation}
R_{\alpha \beta \mu \nu }R^{\alpha \beta \mu \nu }=\frac{6\left(
2r^{6}+K^{2}l^{4}\right) }{r^{6}l^{4}}  \label{Ricci1}
\end{equation}
which confirms that
\begin{equation}
\lim_{r\longrightarrow 0}R_{\alpha \beta \gamma \delta }R^{\alpha \beta
\gamma \delta }\longrightarrow \infty ,
\end{equation}
and so there is a singularity at $r=0$. Also, the Kretschmann scalar is $%
\frac{12}{l^{4}}$ for $r\longrightarrow \infty $ and so asymptotic behavior
of the mentioned spacetime is similar to charged adS BTZ black holes. On the
other hand, comparing the solution (\ref{g(r)pure}) with the solution (\ref%
{g(r)F(R)}) that shows that these solutions are the same provided $K=$ $-%
\frac{\left( 2\mathcal{Q}^{2}\right) ^{\frac{3}{4}}}{2\left(
1+f^{\prime }(R_{0})\right) }$ and $l^{2}=\frac{-6}{R_{0}^{2}}$.
In other words, one can extract charged solution of pure gravity
provided the parameter of the solution (\ref{g(r)pure}) is chosen
suitably. Also we plot $g(r)$ versus $r$ in Fig. \ref{Fig3}, and
find that the presented solutions may be interpreted as black hole
solutions with two horizons, extreme black hole or naked
singularity.

%%%%%%%%%%%%%%%%%%%%%%%%%%%%%%%%%%%%%%%%%%%%%%%%%%%%%%%%%%%%%%%%%%%%%%%%%%%%%%%%%%%%%%
%%%%%%%%%%%%%%%%%%%%%%%%%%%%%%%%%%%%%%%%%%%%%%%%%%%%%%%%%%%%%%%%%%%%%%%%%%%%%%%%%%%%%%

Choosing an $F(R)$ model, we should discuss about its stability. It was
showed that there is no stable ground state for $F(R)$ models if $F(R)\neq 0$
and $F_{R}=0$ \cite{Schmidt}. For the obtained solutions, we find that $%
F(R)=F_{R}=0$. Furthermore, it has been shown that $F_{RR}=d^{2}F/dR^{2}$ is
related to the the effective mass of the dynamical field of the Ricci scalar
\cite{Dolgov}. Therefore, the positive effective mass, a requirement usually
referred to as the Dolgov-Kawasaki stability criterion, leads to the stable
dynamical field \cite{Faraoni}. In order to check the Dolgov-Kawasaki
stability criterion, we should calculate the second derivative of the $F(R)$
function with respect to the Ricci scalar for this specific model%
\begin{equation}
F_{RR}=-\frac{R(n-1)\left( \xi ^{2}+nR^{-2}(1+\xi R)\right) }{n+\xi R},
\label{FRR}
\end{equation}

\begin{figure}[tbp]
\epsfxsize=8cm \centerline{\epsffile{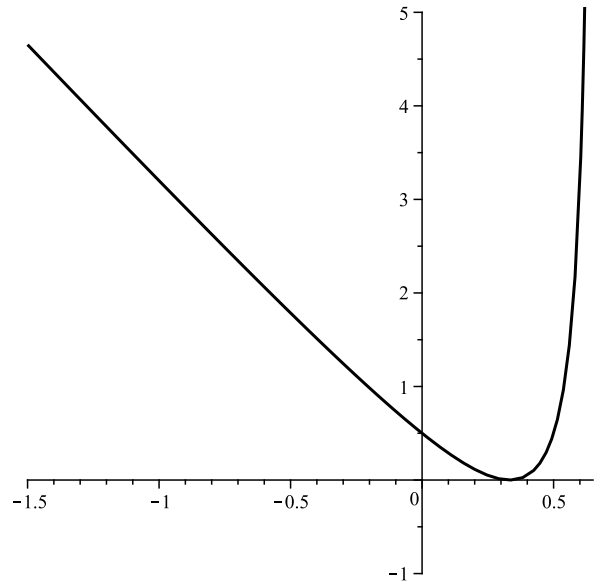}} \caption{ $FRR$
versus $\protect\xi $ for $l=1$, and $n=4$.} \label{Fig4}
\end{figure}

The positivity of $F_{RR}$ depends on the model parameters. The sign of Eq. (%
\ref{FRR}) is not clear for arbitrary values of parameter. Therefore we plot
$F_{RR}$ in Fig. \ref{Fig4} to analyze the sign of $F_{RR}$. This figure
shows that one may obtain stable solution for special values of $\xi $ for
negative cosmological constant. In other words, we can set free parameters
to obtain stable model.

%%%%%%%%%%%%%%%%%%%%%%%%%%%%%%%%%%%%%%%%%%%%%%%%%%%%%%%%%%%%%%%%%%%%%%%%%%%%%%%%%%%%%%%%%%%%%%%%%
%%%%%%%%%%%%%%%%%%%%%%%%%%%%%%%%%%%%%%%%%%%%%%%%%%%%%%%%%%%%%%%%%%%%%%%%%%%%%%%%%%%%%%%%%%%%%%%%%

\section{Thermodynamics}

In this section, we calculate the conserved and thermodynamics quantities of
the black hole solutions in $(2+1)$-dimension for checking the first law of
thermodynamics.

\subsection{\emph{Einstein-PMI and Einstein-CIM gravities}}

The Hawking temperature of the black hole on the outer horizon $r_{+}$, may
be obtained through the use of the definition of surface gravity,
\begin{equation}
T_{+}=\frac{1}{2\pi }\sqrt{-\frac{1}{2}\left( \nabla _{\mu }\chi _{\nu
}\right) \left( \nabla ^{\mu }\chi ^{\nu }\right) },  \label{TemPMI1}
\end{equation}%
where $\chi =\partial /\partial t$ is the Killing vector. We calculate the
Hawking temperature for the PMI black hole solutions, with the following
form
\begin{equation}
T_{+}=\left\{
\begin{array}{c}
\frac{1}{2\pi l^{2}}\left( r_{+}-\frac{q^{2}l^{2}}{r_{+}}\right) \text{ \ \
\ \ \ \ \ }s=1 \\
\frac{1}{4\pi l^{2}}\left( 2r_{+}+\frac{l^{2}\left( 2s-1\right) \left( \frac{%
8q^{2}\left( s-1\right) ^{2}}{\left( 2s-1\right) ^{2}}\right) ^{s}}{r_{+}^{%
\frac{1}{1-2s}}}\right) \text{ \ }otherwise%
\end{array}%
\right. .  \label{TemPMI2}
\end{equation}

Moreover, we known that the electric potential $U$, measured at infinity
with respect to the horizon, is defined by \cite{Cvetic}
\begin{equation}
U=A_{\mu }\chi ^{\mu }\left\vert _{r\rightarrow \infty }\right. -A_{\mu
}\chi ^{\mu }\left\vert _{r=r_{+}}\right. =\left\{
\begin{array}{c}
-q\ln (\frac{r_{+}}{l})\text{ \ \ \ \ \ \ \ \ \ \ \ }s=1 \\
qr_{+}^{\left( \frac{2\left( s-1\right) }{2s-1}\right) }\text{ \ \
\ \ \ \
\ \ }otherwise%
\end{array}%
\right. .  \label{ElecPoPMI1}
\end{equation}

Usually entropy of the black holes satisfies the so-called area law of
entropy which states that the black hole entropy equals to one-quarter of
horizon area \cite{Beckenstein}. Since the area law of the entropy is
universal, and applies to all kinds of black holes in Einstein gravity \cite%
{Beckenstein}, therefore the entropy of the black holes in $(2+1)$-dimension
is equal to one-quarter of the area of the horizon, i.e.,
\begin{equation}
S=\frac{\pi r_{+}}{2}.  \label{entropyPMI}
\end{equation}

In order to calculate the electric charge of the black holes, $Q$, one can
obtain the flux of the electromagnetic field at infinity, yielding
\begin{equation}
Q=2^{s-2}\left( \frac{2q\left( s-1\right) }{\left( 2s-1\right) }\right)
^{2s-1}.  \label{QPMI}
\end{equation}

The present spacetime (\ref{Metric}), have boundaries with timelike $%
\left(\xi =\partial /\partial t\right) $ Killing vector field. It is
straightforward to show that for the quasi local mass we have
\begin{equation}
M=\frac{m}{8}.  \label{massPMI}
\end{equation}

Having conserved and thermodynamic quantities at hand, we are in a position
to check the first law of thermodynamics for our solutions. We obtain the
mass as a function of the extensive quantities $S$ and $Q$. One may then
regard the parameters $S$, and $Q$ as a complete set of extensive parameters
for the mass $M(S,Q)$%
\begin{equation}
M(S,Q)=\frac{2^{\frac{2-3s}{2s-1}}\left( 2s-1\right) ^{2}\left( \frac{S}{\pi
}\right) ^{\frac{2\left( s-1\right) }{2s-1}}Q^{\frac{2s}{2s-1}}+\frac{\left(
s-1\right) }{2l^{2}}\left( \frac{S}{\pi }\right) ^{2}}{\left( s-1\right) },
\end{equation}%
we define the intensive parameters conjugate to $S$ and $Q$. These
quantities are the temperature and the electric potential%
\begin{equation}
T=\left( \frac{\partial M}{\partial S}\right) _{Q}\ \ \ \ ,\ \ \ \ U=\left(
\frac{\partial M}{\partial Q}\right) _{S}.  \label{TU}
\end{equation}

Equations (\ref{TU}) are coincide with Eqs. (\ref{TemPMI2}) and
(\ref{QPMI}) and therefore we find that these thermodynamics
quantities satisfy the first law of black hole thermodynamics
\begin{equation}
dM=TdS+UdQ.
\end{equation}

\subsection{\emph{F(R)-CIM gravity}}

Here, we calculate the Hawking temperature for the black hole solutions of $%
F(R)$-$CIM$ gravity. Such as previous method, we use the definition of
surface gravity to calculate the Hawking temperature. So using Eq. (\ref%
{TemPMI1}) and Eq. (\ref{g(r)F(R)}), we have%
\begin{equation}
T_{+}=\frac{r_{+}}{8\pi (1+f_{R})}\left( \frac{\left( 2\mathcal{Q}%
^{2}\right) ^{\frac{3}{4}}}{r_{+}^{3}}-\frac{2}{3}R_{0}\left( 1+f_{R}\right)
\right) .  \label{TemF(R)CPMI}
\end{equation}

The electric charge of the conformally invariant $F(R)$ black hole can be
obtained as%
\begin{equation}
Q=\frac{3\mathcal{Q}^{\frac{1}{2}}}{2^{\frac{13}{4}}}.
\end{equation}

The electric potential $U$ is \cite{Cvetic}%
\begin{equation}
U=A_{\mu }\chi ^{\mu }\left\vert _{r\rightarrow \infty }\right. -A_{\mu
}\chi ^{\mu }\left\vert _{r=r_{+}}\right. =\frac{\mathcal{Q}}{r_{+}}.
\label{elcpoF(R)CPMI}
\end{equation}

By using the Noether charge method for evaluating the entropy associated
with black-hole solutions in $F(R)=R+f(R)$ theory, it was shown that the
area law does not hold, and one obtains a modification of the area law \cite%
{Cognola2005}
\begin{equation}
S=\frac{A}{4}F_{R},
\end{equation}%
where $A$ is the horizon area. Hence, we can write%
\begin{equation}
S=\frac{\pi r_{+}}{2}\left( 1+f_{R}\right) ,
\end{equation}%
which shows that the area law does not hold for the black hole solutions in $%
R+f(R)$ gravity.

Using the same approach, we find that the finite mass can be given by%
\begin{equation}
M=\frac{m}{8}\left( 1+f_{R}\right) .
\end{equation}

Having the conserved and thermodynamic quantities of the system at hand, we
can check the validity of the first law of black-hole thermodynamics. It is
a matter of straightforward calculation to show that the conserved and
thermodynamics quantities satisfy the first law of thermodynamics
\begin{equation}
dM=TdS+UdQ.
\end{equation}

\subsection{\emph{Pure gravity}}

At first, we investigate the Hawking temperature of the pure black hole on
the outer horizon $r_{+}$. It is a matter of calculation to show that
\begin{equation}
T_{+}=\frac{1}{4\pi l^{2}}\left( 2r_{+}-\frac{Kl^{2}}{r_{+}^{2}}\right) ,
\end{equation}%
which shows that the temperature of the solution is positive definite
provided $K>2r_{+}^{3}/l^{2}$.

For the obtained solutions of pure gravity, $F(R)=F_{R}=0$ and therefore
using the method of Ref. \cite{Cognola2005}, one may encounter with zero
entropy. In order to obtain correct nonzero entropy, one may consider the
first law as a fundamental principle, i.e.
\begin{equation}
dS=\frac{1}{T}dM.
\end{equation}

Using the $\partial /\partial t$ as a Killing vector, one can obtain the
following relation for the finite mass
\begin{equation}
M=\frac{m}{8}=\frac{1}{8}\left[ \frac{r_{+}^{2}}{l^{2}}+\frac{K}{r_{+}}%
\right] .
\end{equation}

Hence, we can write
\begin{equation}
dM=\frac{1}{8}\left[
\frac{2r_{+}}{l^{2}}-\frac{K}{r_{+}^{2}}\right] dr_{+},
\end{equation}
to calculate the entropy as follows
\begin{equation}
S=\int \frac{1}{T}dM=\int \frac{1}{\frac{-1}{4\pi }\left( \frac{K}{r_{+}^{2}}
-\frac{2r_{+}}{l^{2}}\right) }\frac{\frac{2r_{+}}{l^{2}}-\frac{K}{r_{+}^{2}}
}{8}dr_{+},
\end{equation}
\begin{equation}
S=\frac{\pi }{2}\int dr_{+}=\frac{\pi r_{+}}{2},
\end{equation}
which is nothing but the are a law for the entropy. In other words,
considering the $F(R)$ gravity with $F_R=0$, one may use the the so-called
area law instead of modified area law \cite{Cognola2005}.

\section{Conclusions}

In this paper, we regarded some various gravitational theories in $(2+1)$
dimensions. At first, we reviewed the Einstein-Maxwell gravity and then we
considered Einstein-PMI and Einstein-CIM theories. We obtained the black
hole solutions of these theories and studied their properties. We showed
that the black holes of the mentioned nonlinear electrodynamics covered with
one horizon such as uncharged (Schwarzschild) black holes.

After that, we took into account the pure $F(R)$ gravity as well as $F(R)$%
-CIM theory to obtain the black hole solutions. We showed that for the pure
gravity, one can extract the electrical charge and cosmological constant,
simultaneously. In addition, we showed that the asymptotic behavior of
obtained solutions is adS.

Finally, we computed the conserved and thermodynamics quantities of the
black hole solutions in $(2+1)$-dimension for various theories and found
that they satisfy the first law of thermodynamics. In other words, regarding
the pure $F(R)$ gravity, we considered the first law as a fundamental
principle and proposed that despite the modified area law of entropy ($S=%
\frac{A}{4}F_{R}$) being true for $F_{R} \neq 0$ model, but we should regard
the area law for $F_{R}=0$ cases.

In this paper, we only considered static three dimensional spacetime.
Therefore it is worthwhile to think about the cosmological scenario by
regarding time dependent spacetime with exact solutions of equations of
motion depending on time. One more subject of interest for further study is
finding the corresponding scalar-tensor gravity solutions and these subjects
are under examination.

\begin{acknowledgements}
B. Eslam Panah acknowledges S. Panahiyan and A. Sheykhi for
helpful discussions. S. H. Hendi and B. Eslam Panah thank Shiraz
University Research Council. This work has been supported
financially by the Research Institute for Astronomy and
Astrophysics of Maragha, Iran.
\end{acknowledgements}

\end{document}